# *Ab initio* calculation of atomic solid hydrogen phases based on Gutzwiller many-body wave functions

Zhuo Ye,[1,*] Jun Liu,[1] Yong-Xin Yao,[1,2] Feng Zhang,[1,2] Kai-Ming Ho,[2] Cai-Zhuang Wang,[1,2,*]

[1] Ames National Laboratory – US DOE, Ames, Iowa 50011, United States

[2] Department of Physics and Astronomy, Iowa State University, Ames, Iowa 50011, United States

* E-mail: zye@iastate.edu (Z. Ye), wangcz@ameslab.gov (C. Z. Wang)

ABSTRACT

We apply two *ab initio* many-body methods based on Gutzwiller wave functions, i.e., correlation matrix renormalization theory (CMRT) and Gutzwiller conjugate gradient minimization (GCGM), to the study of crystalline phases of atomic hydrogen. Both methods avoid empirical Hubbard $U$ parameters and are free from double-counting issues. CMRT employs a Gutzwiller-type approximation that enables efficient calculations, while GCGM goes beyond this approximation to achieve higher accuracy at higher computational cost. By benchmarking against available quantum Monte Carlo results, we demonstrate that while both methods are more accurate than the widely used density-functional theory (DFT), GCGM systematically captures additional correlation energy missing in CMRT, leading to significantly improved total energy predictions. We also show that by including the correlation energy $E_c$ from LDA in the CMRT calculation, CMRT+$E_c$ produces energy in better agreement with the QMC results in these hydrogen lattice systems.

1. INTRODUCTION

When the Schrödinger equation was introduced into quantum mechanics in 1926, there were hopes that it would soon enable a complete understanding and description of chemistry, allowing for accurate predictions of materials and rendering many experimental methods unnecessary. As Pauli noted, "the whole of chemistry are thus completely known, and the difficulty is only that the exact application of these laws leads to equations much too complicated to be soluble" [1]. Nearly a century later, this statement remains largely true. It is still a formidable challenge to obtain exact solutions of the Schrödinger equation for most molecular systems, not to mention condensed bulk systems. Introducing approximations is necessary in order to solve the Schrödinger equation, and selecting the most suitable approximation for a given problem is one of the central challenges in quantum chemistry and condensed matter physics.

One commonly used approximation is to replace the Coulomb interactions between electrons by some mean-field potentials, i.e., the one-electron approximation. A standard approach using this approximation is the Hartree-Fock (HF) method [2-4]. It is the conceptual foundation for a bunch of post-Hartree-Fock methods, which seek to capture the correlation effects by different higher-level methods, such as configuration interaction (CI) or truncated CI [5-7], many-body perturbation theory [8,9], and coupled-cluster theory [8,10,11]. Nevertheless, these post-HF methods aimed at addressing the electron-electron correlations neglected in the HF approximation are often expensive or prohibitive when applying to bulk materials.

An alternative approximation is the one used in Kohn-Sham density functional theory (DFT) [12,13], which is the most widely used computational method for electronic structure calculations in molecules and condensed matter systems today. DFT achieves high efficiency by using the electron density, rather than the many-body wave function, as the central variational quantity, and by mapping the interacting electron problem onto an auxiliary non-interacting system. In this framework, the total energy is decomposed into kinetic energy, classical Hartree energy, and exchange-correlation (XC) energy, the latter of which can only be treated using approximate functionals. DFT generally performs well when the system of interest resembles the reference system used in the construction of the XC energy functional, such as the homogeneous electron gas. However, in strongly correlated materials, where electron-electron interactions dominate the behavior of the materials, conventional XC energy functionals often fail to capture the essential many-body effects. Hybrid approaches that merge DFT with many-body techniques have been developed with improved accuracy, such as DFT + onsite Coulomb interaction (DFT + U) [14,15] or DFT + Gutzwiller [16-20], yet the accuracy of these methods often ties to the choice of the screened Coulomb parameters. It is highly desirable to develop *ab initio* many-body approaches that are computationally affordable and with reasonable accuracy to study strongly correlated electron materials.

Moving towards this goal, we have been developing methods based on Gutzwiller variational wave function (GWF) [21-23], namely, the correlation matrix renormalization theory (CMRT) [24-28] and the Gutzwiller conjugate gradient minimization (GCGM) method [29-33]. The two methods use different types of approximation, resulting in different levels of speed and accuracy. The GWF is constructed by applying the Gutzwiller projector on a trial non-interacting wave function so that each on-site valence electronic configuration is assigned a variable amplitude and phase factor. The reason we chose this scheme is that the GWF, when used together with the Gutzwiller approximation, can reach a high computational efficiency comparable to HF. In the Gutzwiller approximation, an element of one-particle density matrix expressed in GWF $|\Psi_{GWF}\rangle$ can be approximated by the multiplication of the corresponding element in the trial non-interacting wave-function $|\Psi_0\rangle$ and the one-electron renormalization z-factor that is assigned to each of the two relevant sites $i, j$ : $\langle c_{i\alpha}^\dagger c_{j\beta} \rangle_{GWF} \approx z_{i\alpha}^{GA} z_{j\beta}^{GA} \langle c_{i\alpha}^\dagger c_{j\beta} \rangle_0$, where $z_{i\alpha}^{GA}$ and $z_{j\beta}^{GA}$ are the Gutzwiller renormalization factors on site $i$ and $j$ for spin-orbital $\alpha$ and $\beta$, respectively. $\langle \hat{O} \rangle_{GWF}$ is a short-hand notation for $\langle \Psi_{GWF} | \hat{O} | \Psi_{GWF} \rangle$, and $\langle \hat{O} \rangle_0$ for $\langle \Psi_0 | \hat{O} | \Psi_0 \rangle$ with any operator $\hat{O}$. Evaluation of the density matrix with the Gutzwiller approximation can be very efficient. However, the Gutzwiller approximation recovers the exact GWF results only in the infinite dimension limit. It could be a major source of inaccuracy in finite dimensions. By comparing the total energy expression with the exact solution for the H$_2$ molecule under a minimal basis set, we found that the one-electron renormalization factor $z_{i\alpha}$ is better expressed as the square root of $z_{i\alpha}^{GA}$: $\langle c_{i\alpha}^{\dagger} c_{j\beta} \rangle_{GWF} \approx z_{i\alpha} z_{j\beta} \langle c_{i\alpha}^{\dagger} c_{j\beta} \rangle_0$ with $z_{i\alpha} = \sqrt{z_{i\alpha}^{GA}}$. This refined Gutzwiller-type approximation is adopted in CMRT and has been demonstrated to be capable of improving the accuracy for most of the systems we have tested [24]. Nevertheless, this type of approximation manifestly "decouples" the two correlated sites $i, j$ by using a site-wise factorization, rendering a seemingly lack of pair-environment dependence. As a result, discrepancies were still observed when compared to benchmark data. The GCGM method was then introduced to further improve accuracy by using a more rigorous way to evaluate the density matrix without relying on any Gutzwiller-type approximations. Another advantage of GCGM is that it overcomes some limitations of the GWF itself. The Gutzwiller projector introduces correlations into $|\Psi_0\rangle$ only via an on-site correlation factor. In the case that some inter-site correlation is absent in $|\Psi_0\rangle$, it will also be missing in the GWF. In the GCGM, a site-site constraint is applied in the GWF to compensate for such absent inter-site correlation [31]. In general, GCGM is more accurate than CMRT with the above two improvements, but at the cost of extra computational effort.

Our previous work using the GCGM has demonstrated improved accuracy with benchmark tests on molecules and model systems like one- and two-band Hubbard models [29,30,33]. As we aim to eventually develop a numerical tool to access realistic condensed matter systems, the next benchmark test is naturally one of the simplest realistic bulk systems. In this work, we choose solid atomic hydrogen phases as our first three-dimensional periodic system to benchmark GCGM and compare the results with the reference data given by quantum Monte Carlo (QMC) [34-37]. At the same time, we present the CMRT results to show the difference between the two GWF-based methods. Our results indicate that GCGM captures the extra correlation energy missing from CMRT and yields a more accurate energy calculation. By adopting the local density approximation (LDA) for the correlation energy $E_c$ and including it in the CMRT calculation, CMRT+$E_c$ produces energy in better agreement with the QMC results in these hydrogen lattice systems.

## 2. METHODS AND FORMALISM

For a periodic system, the general nonrelativistic Hamiltonian can be expressed as,

$$\hat{H} = \sum_{Ii\alpha,Jj\beta,\sigma} t_{Ii\alpha,Jj\beta} c^\dagger_{Ii\alpha\sigma} c_{Jj\beta\sigma} + \frac{1}{2} \sum_{\substack{Ii\alpha,Jj\beta \\ Kk\gamma,Ll\delta,\sigma\sigma'}} u(Ii\alpha,Jj\beta;Kk\gamma,Ll\delta) c^\dagger_{Ii\alpha\sigma} c^\dagger_{Jj\beta\sigma'} c_{Ll\delta\sigma'} c_{Kk\gamma\sigma}, \quad (1)$$

with $I,J,K,L$ indices the unit cells, $i,j,k,l$ the atomic sites, $\alpha,\beta,\gamma,\delta$ the orbitals, and $\sigma,\sigma'$ the spin. The one-electron hopping integral, $t$, and the two-electron Coulomb integral, $u$, are defined as,

$$t_{Ii\alpha,Jj\beta} = \left\langle \phi_{Ii\alpha} \middle| \hat{T} + \hat{V}_{ion} \middle| \phi_{Jj\beta} \right\rangle, \quad (2)$$

and

$$u(Ii\alpha,Jj\beta;Kk\gamma,Ll\delta) = \iint d\mathbf{r} d\mathbf{r}' \phi^*_{Ii\alpha}(\mathbf{r}) \phi^*_{Jj\beta}(\mathbf{r}') \hat{U}(\mathbf{r}-\mathbf{r}') \phi_{Ll\delta}(\mathbf{r}') \phi_{Kk\gamma}(\mathbf{r}), \quad (3)$$

where $\hat{T}$, $\hat{V}_{ion}$, and $\hat{U}$ are the operators for the kinetic energy, the ion-electron interaction, and the Coulomb interaction, respectively. $\phi_{Ii\alpha}$ is the basis orbital $\alpha$ at atomic site $Ii$. We consider a trial wave function $|\Psi_0\rangle$ that is often non-interacting, i.e., a single Slater determinant, which can be written as the projection onto site-wise Fock state configurations,

$$|\Psi_0\rangle = \sum_{\{\Gamma_{Ii}\}} |\{\Gamma_{Ii}\}\rangle \langle\{\Gamma_{Ii}\}|\Psi_0\rangle = \sum_{\{\Gamma_{Ii}\}} \lambda_{\{\Gamma_{Ii}\}} |\{\Gamma_{Ii}\}\rangle, \quad (4)$$

where the summation runs through all possible single-atom configurations $\{\Gamma_{Ii}\}$. The coefficient $\lambda_{\{\Gamma_{Ii}\}}$ is defined as $\lambda_{\{\Gamma_{Ii}\}} = \langle\{\Gamma_{Ii}\}|\Psi_0\rangle$ and $\Gamma_{Ii}$ is the onsite configuration at atomic site $Ii$, which is defined as a Fock state $|\Gamma_{Ii}\rangle \equiv \prod_{\alpha\sigma \in \Gamma_{Ii}} c^\dagger_{\alpha\sigma} |\varnothing\rangle$. Here the creation operator $c^\dagger_{\alpha\sigma}$ creates an electron at the orbital-$\alpha$ with spin-$\sigma$ in the vacuum state $|\varnothing\rangle$. The GWF is constructed by applying the Gutzwiller projector $\hat{G}$ on $|\Psi_0\rangle$,

$$|\Psi_{GWF}\rangle = \hat{G}|\Psi_0\rangle = \prod_{Ii}\left(\sum_{\Gamma} g(\Gamma_{Ii})|\Gamma_{Ii}\rangle\langle\Gamma_{Ii}|\right)|\Psi_0\rangle = \sum_{\{\Gamma_{Ii}\}}\left(\prod_{Ii} g(\Gamma_{Ii})\right)\lambda_{\{\Gamma_{Ii}\}}|\{\Gamma_{Ii}\}\rangle, \quad (5)$$

where $g(\Gamma_{Ii})$ is the Gutzwiller variational parameter determining the occupation probability of the on-site configuration $|\Gamma_{Ii}\rangle$. Due to translational invariance, $g(\Gamma_{Ii})$ is independent of unit cell thus $g(\Gamma_{Ii}) = g(\Gamma_i)$. The total energy of the system can be expressed in terms of the one-particle density matrix (1PDM) and the two-particle correlation matrix (2PCM),

$$E_{GWF} = \sum_{Ii\alpha, Jj\beta, \sigma} t_{Ii\alpha, Jj\beta} \langle c^\dagger_{Ii\alpha\sigma} c_{Jj\beta\sigma}\rangle_{GWF} + \frac{1}{2}\sum_{\substack{Ii\alpha, Jj\beta \\ Kk\gamma, Ll\delta, \sigma\sigma'}} u(Ii\alpha, Jj\beta; Kk\gamma, Ll\delta)\langle c^\dagger_{Ii\alpha\sigma} c^\dagger_{Jj\beta\sigma'} c_{Ll\delta\sigma'} c_{Kk\gamma\sigma}\rangle_{GWF}.$$

(6)

Then we evaluate the on-site 2PCM $\langle c^\dagger_{Ii\alpha\sigma} c^\dagger_{Ii\beta\sigma'} c_{Ii\delta\sigma'} c_{Ii\gamma\sigma}\rangle_{GWF}$ rigorously and use Wick's theorem [26,38] to evaluate the inter-site two-particle correlation matrix approximately:

$$\langle c^\dagger_{Ii\alpha\sigma} c^\dagger_{Jj\beta\sigma'} c_{Ll\delta\sigma'} c_{Kk\gamma\sigma}\rangle_{GWF} \approx$$
$$\langle c^\dagger_{Ii\alpha\sigma} c_{Kk\gamma\sigma}\rangle_{GWF} \langle c^\dagger_{Jj\beta\sigma'} c_{Ll\delta\sigma'}\rangle_{GWF} - \delta_{\sigma\sigma'}\langle c^\dagger_{Ii\alpha\sigma} c_{Ll\delta\sigma}\rangle_{GWF} \langle c^\dagger_{Jj\beta\sigma} c_{Kk\gamma\sigma}\rangle_{GWF}.$$

(7)

The total energy can be expressed as,

$$E_{GWF} \approx \sum_{Ii\alpha, Jj\beta, \sigma} t_{Ii\alpha, Jj\beta} \langle c^\dagger_{Ii\alpha\sigma} c_{Jj\beta\sigma}\rangle_{GWF} + \frac{1}{2}\sum_{\substack{Ii, \alpha\beta\gamma\delta \\ \sigma\sigma'}} u(Ii\alpha, Ii\beta; Ii\gamma, Ii\delta)\langle c^\dagger_{Ii\alpha\sigma} c^\dagger_{Ii\beta\sigma'} c_{Ii\delta\sigma'} c_{Ii\gamma\sigma}\rangle_{GWF} +$$
$$\frac{1}{2}\sum'_{\substack{Ii\alpha, Jj\beta \\ Kk\gamma, Ll\delta, \sigma\sigma'}} \left(u(Ii\alpha, Jj\beta; Kk\gamma, Ll\delta) - \delta_{\sigma\sigma'} u(Ii\alpha, Jj\beta; Ll\delta, Kk\gamma)\right) \langle c^\dagger_{Ii\alpha\sigma} c_{Kk\gamma\sigma}\rangle_{GWF} \langle c^\dagger_{Jj\beta\sigma'} c_{Ll\delta\sigma'}\rangle_{GWF},$$

(8)

where $\Sigma'$ indicates that the on-site terms ($Ii = Jj = Kk = Ll$) are excluded from the summation.

The use of Wick's theorem will result in errors from the HF-type factorization, which can be reduced by using a sum-rule correction that shifts non-trivial inter-site two-particle terms into one-particle and onsite two-particle terms that will be evaluated in a more rigorous way. The sum-rule term is constructed on charge conservation and is included in the original Hamiltonian as [26,27],

$$\hat{H} \to \hat{H} + \hat{H}_{sr} = \hat{H} + \frac{1}{2}\sum_{Ii\alpha}\lambda_{i\alpha}\left(\hat{n}_{Ii\alpha\sigma}\left(\sum_{Jj\beta\sigma'}\hat{n}_{Jj\beta\sigma'} - N_e\right)\right), \tag{9}$$

where $N_e$ is the total number of electrons in the system, $\hat{n}_{i\alpha\sigma}$ the electron number operator, $\hat{n}_{Ii\alpha\sigma} = c^{\dagger}_{Ii\alpha\sigma}c_{Ii\alpha\sigma}$, and $\lambda_{i\alpha}$ the weighted average of the relevant inter-site 2-electron Coulomb integrals,

$$\lambda_{i\alpha} = \frac{\sum_{I \neq J, j\beta\sigma\sigma'} u(Ii\alpha, Jj\beta; Jj\beta, Ii\alpha)\left|\left\langle c^{\dagger}_{Ii\alpha\sigma}c_{Jj\beta\sigma'}\right\rangle_{GWF}\right|^4}{\sum_{I \neq J, j\beta\sigma\sigma'}\left|\left\langle c^{\dagger}_{Ii\alpha\sigma}c_{Jj\beta\sigma'}\right\rangle_{GWF}\right|^4}. \tag{10}$$

We note that the sum-rule term is included in the Hamiltonian for both CMRT and GCGM methods, and so the two methods share the same Hamiltonian. Nevertheless, the two approaches utilize different ways to evaluate the one-particle density matrix.

In CMRT, the one-particle density matrix is evaluated approximately as,

$$\left\langle c^{\dagger}_{Ii\alpha\sigma}c_{Jj\beta\sigma}\right\rangle_{GWF} = \begin{cases} \left\langle c^{\dagger}_{Ii\alpha\sigma}c_{Ii\alpha\sigma}\right\rangle_0 & \text{for } Ii\alpha = Jj\beta \\ z_{i\alpha\sigma}z_{j\beta\sigma}\left\langle c^{\dagger}_{Ii\alpha\sigma}c_{Jj\beta\sigma}\right\rangle_0 & \text{for } Ii\alpha \neq Jj\beta, \end{cases} \tag{11}$$

where $z_{i\alpha\sigma} = f\left(z^{GA}_{i\alpha\sigma}\right)$ is a function of the Gutzwiller renormalization factor $z^{GA}_{i\alpha\sigma}$ that is determined by the Fock states occupation probability [26], and $\left\langle c^{\dagger}_{Ii\alpha\sigma}c_{Jj\beta\sigma}\right\rangle_0$ the density matrix for the non-interacting trial wavefunction. As explained in the introduction, CMRT uses a functional form of $f(z^{GA}_{i\alpha\sigma}) = \sqrt{z^{GA}_{i\alpha\sigma}}$ to define the one-particle density matrix in its Gutzwiller-type approximation. This enables an efficient evaluation of the density matrix from Eq. (11) since the non-interacting density matrix is readily available, endowing CMRT with a computational speed comparable to HF calculation. However, the approximation indicated in Eq. (11) has several limitations. Firstly, it requires $\left\langle c^{\dagger}_{Ii\alpha\sigma}c_{Ii\alpha\sigma}\right\rangle_{GWF} = \left\langle c^{\dagger}_{Ii\alpha\sigma}c_{Ii\alpha\sigma}\right\rangle_0 = n^0_{Ii\alpha\sigma}$. In other words, if the initial non-interacting wave function gives an orbital occupancy far away from the exact solution, it is going to be technically involved to bring back the correct local physics. Secondly, as shown in the expression $z_{i\alpha\sigma}z_{j\beta\sigma}$, this approximation "decouples" the two atomic sites and may not be able to accurately capture the inter-site correlation. Due to these limitations, Eq. (11) sometimes yields unsatisfactory results.

In order to achieve more accurate results, we developed GCGM and proposed a more rigorous way to estimate the density matrix,

$$\left\langle c_{Ii\alpha\sigma}^{\dagger} c_{Ii\beta\sigma} \right\rangle_{GWF} = \frac{1}{\left\langle \Psi_{GWF} | \Psi_{GWF} \right\rangle_{Ii,Jj}} \sum_{\Gamma_{Ii},\Gamma'_{Ii},\Gamma_{Jj}} \left\langle \Gamma_{Ii} | c_{Ii\alpha\sigma}^{\dagger} c_{Ii\beta\sigma} | \Gamma'_{Ii} \right\rangle \cdot g(\Gamma_i) g(\Gamma'_i) g(\Gamma_j)^2 \xi_{\Gamma_{Ii},\Gamma_{Jj},\Gamma'_{Ii},\Gamma_{Jj}},$$
(12)

$$\left\langle c_{Ii\alpha\sigma}^{\dagger} c_{Jj\beta\sigma} \right\rangle_{GWF} = \frac{1}{\left\langle \Psi_{GWF} | \Psi_{GWF} \right\rangle_{Ii,Jj}} \sum_{\Gamma_{Ii},\Gamma_{Jj},\Gamma'_{Ii},\Gamma'_{Jj}} \left\langle \Gamma_{Ii} | c_{Ii\alpha\sigma}^{\dagger} | \Gamma'_{Ii} \right\rangle \left\langle \Gamma_{Jj} | c_{Jj\beta\sigma} | \Gamma'_{Jj} \right\rangle \cdot g(\Gamma_i) g(\Gamma_j) g(\Gamma'_i) g(\Gamma'_j) \xi_{\Gamma_{Ii},\Gamma_{Jj},\Gamma'_{Ii},\Gamma'_{Jj}} \text{ for } (I,i) \neq (J,j),$$
(13)

where the atomic site $Jj$ is chosen to be the nearest neighbor of site $Ii$ in Eq. (12) and

$$\left\langle \Psi_{GWF} | \Psi_{GWF} \right\rangle_{Ii,Jj} = \sum_{\Gamma_{Ii},\Gamma_{Jj}} \xi_{\Gamma_{Ii},\Gamma_{Jj},\Gamma_{Ii},\Gamma_{Jj}} g(\Gamma_i)^2 g(\Gamma_j)^2.$$
(14)

Here, $\xi_{\Gamma_{Ii},\Gamma_{Jj},\Gamma'_{Ii},\Gamma'_{Jj}}$ are pre-factors determined from the non-interacting $|\Psi_0\rangle$ and Gutzwiller variational parameters,

$$\xi_{\Gamma_{Ii},\Gamma_{Jj},\Gamma'_{Ii},\Gamma'_{Jj}} = \sum_{\{\Gamma_{Kk}, Kk \neq Ii,Jj\}} \prod_{Kk} g(\Gamma_k)^2 \left\langle \Psi_0 | \Gamma_{Ii}, \Gamma_{Jj}, \{\Gamma_{Kk}\} \right\rangle \left\langle \Gamma'_{Ii}, \Gamma'_{Jj}, \{\Gamma_{Kk}\} | \Psi_0 \right\rangle.$$
(15)

The onsite 2-particle correlation matrix can be evaluated similarly as,

$$\left\langle c_{Ii\alpha\sigma}^{\dagger} c_{Ii\beta\sigma'}^{\dagger} c_{Ii\delta\sigma'} c_{Ii\gamma\sigma} \right\rangle_{GWF} = \frac{1}{\left\langle \Psi_{GWF} | \Psi_{GWF} \right\rangle_{Ii,Jj}} \sum_{\Gamma_{Ii},\Gamma'_{Ii},\Gamma_{Jj}} \left\langle \Gamma_{Ii} | c_{Ii\alpha\sigma}^{\dagger} c_{Ii\beta\sigma'}^{\dagger} c_{Ii\delta\sigma'} c_{Ii\gamma\sigma} | \Gamma'_{Ii} \right\rangle \cdot g(\Gamma_i) g(\Gamma'_i) g(\Gamma_j)^2 \xi_{\Gamma_{Ii},\Gamma_{Jj},\Gamma'_{Ii},\Gamma_{Jj}},$$
(16)

where the atomic site $Jj$ is chosen to be the nearest neighbor of site $Ii$. Rigorous evaluation of $\xi_{\Gamma_{Ii},\Gamma_{Jj},\Gamma'_{Ii},\Gamma'_{Jj}}$ from Eq. (15) is formidable, as the computational complexity grows exponentially with respect to the number of atomic sites in the summation. Therefore, we use an approximation to efficiently evaluate $\xi_{\Gamma_{Ii},\Gamma_{Jj},\Gamma'_{Ii},\Gamma'_{Jj}}$ from $\xi^0_{\Gamma_{Ii},\Gamma_{Jj},\Gamma'_{Ii},\Gamma'_{Jj}}$ that is determined from $|\Psi_0\rangle$ only,

$$\xi^0_{\Gamma_{Ii},\Gamma_{Jj},\Gamma'_{Ii},\Gamma'_{Jj}} = \sum_{\{\Gamma_{Kk},Kk\neq Ii,Jj\}} \langle\Psi_0|\Gamma_{Ii},\Gamma_{Jj},\{\Gamma_{Kk}\}\rangle\langle\Gamma'_{Ii},\Gamma'_{Jj},\{\Gamma_{Kk}\}|\Psi_0\rangle. \qquad (17)$$

The detail of the approximation can be found in Refs. [31,33] and will not be repeated here for conciseness.

The total energy $E_{GWF}$ can be estimated from Eq. (8) after $\langle c^\dagger_{Ii\alpha\sigma} c_{Jj\beta\sigma}\rangle_{GWF}$, and $\langle c^\dagger_{Ii\alpha\sigma} c^\dagger_{Ii\beta\sigma'} c_{Ii\sigma'} c_{Ii\sigma}\rangle_{GWF}$ expressed explicitly as a function of $\{g(\Gamma_i)\}$. In GCGM, $E_{GWF}$ is then minimized with respect to $\{g(\Gamma_i)\}$ using the conjugate gradient method with the gradient $\frac{\partial E_{GWF}}{\partial g(\Gamma_i)}$ evaluated analytically. The calculation of the different derivatives $\frac{\partial E_{GWF}}{\partial g(\Gamma_i)}$ is completely decoupled, and thus, is readily parallelizable.

On the other hand, CMRT adopts a similar total energy expression in Eq. 8. However, CMRT does not directly use $\{g(\Gamma_i)\}$ as the variational parameters, but instead, use $z_{i\alpha\sigma}$, the Gutzwiller renormalization factor with $\{g(\Gamma_i)\}$ expressed inside. Additionally, CMRT takes the non-interacting trial wave function as part of the variational parameter set and updates it during the self-consistent charge iteration, similar to the standard variational procedure in DFT. To reach an energy minimum, energy gradients with respect to all these variation parameters must vanish. This procedure brings up a set of highly nonlinear equations to be solved self-consistently with Newton's method for the physical ground state wave function and energy. The electronic correlation energy, $E_c$, can be included in Eq. 8, but is currently not variationally treated. It is evaluated with the optimal parameters solved from the self-consistent equation set and added directly onto the ground state energy.

## 3. RESULTS AND DISCUSSION

In this work, four lattices of atomic solid hydrogen are studied: face-center cubic (FCC), body-center cubic (BCC), simple cubic (SC) and diamond (DIA). The crystal structures of atomic solid hydrogen considered in this benchmark are not intended to represent the thermodynamically stable phases under the corresponding pressure conditions. They are used solely as well-defined periodic test systems for method-to-method comparison. Although the zero-point energy is important in determining the relative phase stability for hydrogen phases [34,35], it is often not critical for systems made of other elements. Since this work serves as a benchmark test for GCGM and CMRT, both of which are developed as tools for studying general correlated systems, we limit our focus to static lattices with fixed atomic sites to ensure generality. Both CMRT and GCGM calculations are interfaced with the Hartree-Fock (HF) module of the Vienna *Ab Initio* Simulation Package (VASP) [39]. The quasi-atomic minimal basis set orbitals (QUAMBOs) [40,41] are used as local basis orbitals with the minimum basis orbitals chosen as $1s$ for hydrogen. The QUAMBOs are constructed from LDA calculations and with a pseudopotential defined with the projector-augmented wave (PAW) method [42] and the Ceperley-Alder exchange-correlation functional [43]. A plane wave basis set was generated in VASP with the default energy cutoff specified by the pseudopotential for hydrogen. The Brillouin zone sampling scheme provided by VASP automatically generates K-point grids for the four types of crystals, maintaining a $R_k$ length of 24 ($R_k = 24$). Specifically, this amounts to a $16 \times 16 \times 16$ uniform mesh at the equilibrium lattice constant of $r = 1.5$ Å for the SC lattice. We benchmark our CMRT and GCGM methods by comparing the equation of state from our calculations with those available in the literature using other methods. In particular, we choose the results given by QMC [44,45] as the reference data for comparison. For the QMC calculation, both variational Monte Carlo and fixed-node diffusion Monte Carlo (DMC) are used in the reference works [34-36]. Here, we only use the results from DMC calculation as reference data. The DMC calculation is firstly performed for finite systems and then is extrapolated to the infinite size limit.

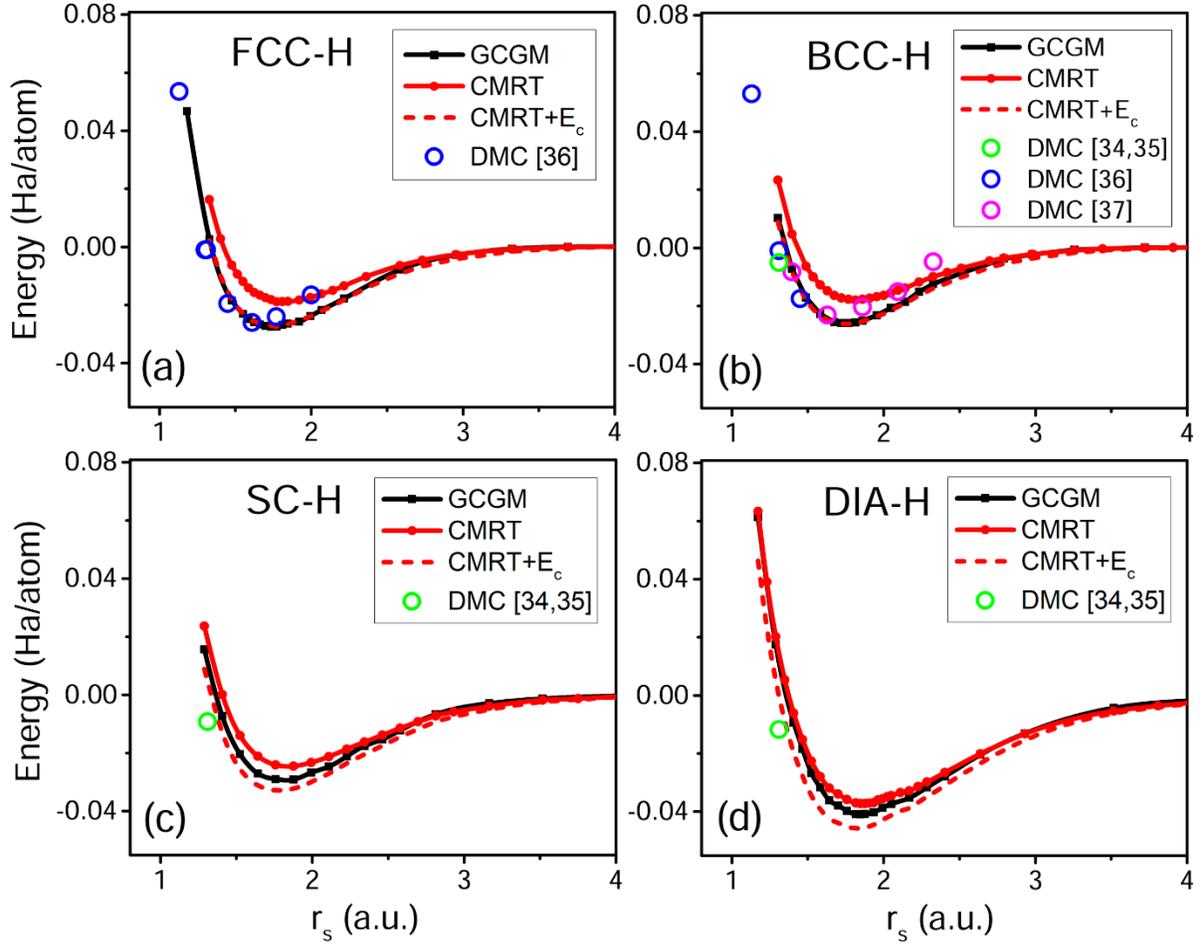

Figure 1. Equation of state for (a) FCC, (b) BCC, (c) SC, and (d) DIA-hydrogen given by GCGM, CMRT, CMRT+ $E_c$ and DMC [34-37].

Figure 1 plots energy per atom as a function of $r_s$ for the 4 atomic solid hydrogen phases obtained by GCGM and CMRT. The Wigner–Seitz radius $r_s$ is the radius of a sphere whose volume is equal to the mean volume per atom in a solid. All energy curves are aligned at the atomic limit, which is approached at around $r_s$ =5 a.u.. The results from DMC [34-37] are also plotted as a reference. For FCC- and BCC-H, the equilibrium energy given by GCGM agrees well with the ones given by DMC, while the equilibrium lattice given by GCGM is a bit larger than the reference results. Overall, the energy curves given by GCGM align reasonably well with the DMC results, with some discrepancies observed at lattice sizes larger than equilibrium ($r_s \sim 2$ a.u.). On the other hand, CMRT underestimates the binding energy when compared to the reference data. However, inclusion of the correlation energy $E_c$ largely improves the CMRT energy, making the energy curve almost overlap with the one given by GCGM. This indicates that CMRT in its current formalism does not fully capture the correlation energy of the valence electrons, likely due to the overly simplified evaluation of the interacting one-particle density matrix, which is a key distinction between GCGM and CMRT. For SC- and DIA-H, there are no DMC results to compare with at equilibrium lattices. At $r_s = 1.3$ a.u., all the 3 methods (GCGM, CMRT and CMRT+$E_c$) underestimate the binding energy compared to the DMC results, with CMRT+$E_c$ providing the closest estimate.

In order to analyze what causes the energy difference between GCGM and CMRT, we plot the two leading components of energy in Fig. 2: the double occupancy $\langle c^\dagger_{Ii\uparrow} c_{Ii\uparrow} c^\dagger_{Ii\downarrow} c_{Ii\downarrow} \rangle_{GWF}$ and the nearest neighbor hopping $\langle c^\dagger_{Ii\sigma} c_{Jj\sigma} \rangle_{GWF}$ with $(Ii, Jj)$ being nearest neighbors. A common pattern can be seen for all the four phases. At smaller $r_s < 1.5$ a.u., the two methods give almost the same nearest neighbor hopping, and double occupancy is the major source of the energy difference between the two methods. Around the equilibrium with $r_s$ being $1.5 \sim 2$ a.u., while double occupancy still contributes to the energy difference, nearest-neighor hopping starts playing a role. At larger lattices $r_s > 2.2$ a.u., nearest-neighor hopping becomes the major source of the energy difference. Finally, as it approaches the atomic limit, the two methods give around the same double occupancy and nearest-neighor hopping. Among the 4 phases, the difference in nearest-neighbor hopping is of a similar scale, while the difference in double occupancy is the smallest for DIA-H. As a result, the energy difference between GCGM and CMRT is also the smallest for DIA-H, as can be seen in Fig. 1.

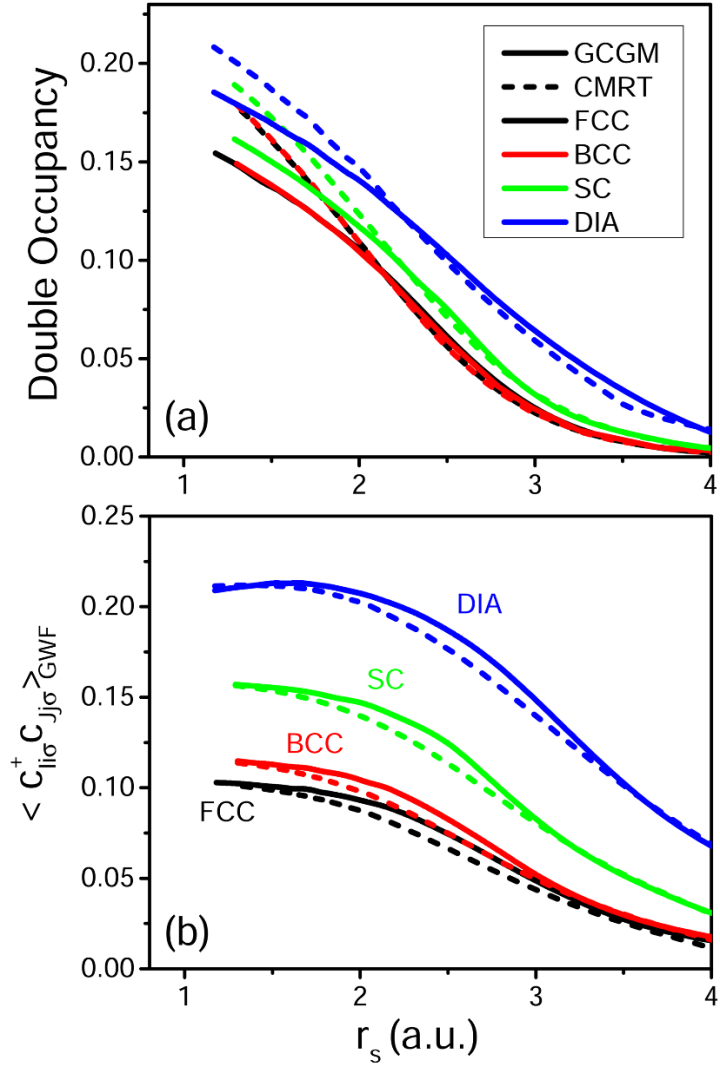

Figure 2. (a) Double occupancy as a function of lattice given by GCGM and CMRT. (b) $\left\langle c^{\dagger}_{Ii\sigma} c_{Jj\sigma} \right\rangle_{GWF}$ with $(Ii, Jj)$ being nearest neighbors, as a function of $r_s$ obtained by GCGM and CMRT.

In Fig. 3, with DMC results being used as a reference, we plot in each subfigure the equation of state for all the 4 phases obtained from GCGM, CMRT and CMRT $+E_c$, respectively. As a comparison, we also plot the results given by the PAW method within DFT as implemented in VASP. In the DFT calculations, the exchange and correlation energy functional is treated within the spin-polarized generalized gradient approximation (GGA) and parameterized by Perdew-Burke-Ernzerhof formula (PBE) [46]. A plane wave basis set was generated in VASP with the default energy cutoff specified by the pseudopotential for hydrogen. The Brillouin zone sampling is automatically generated with $R_k = 24$ as well. The DFT calculation overestimates the binding energy around equilibrium and at $r_s$ bigger than equilibrium while giving energies in very good agreement with DMC at smaller $r_s \sim 1.3$ a.u.. Since the homogeneous electron gas is a standard reference system in constructing approximate exchange–correlation functionals, it is expected that the accuracy of DFT calculations improves with increasing pressure. In contrast, CMRT underestimates the binding energy at equilibrium and at $r_s$ smaller than equilibrium. Overall, GCGM and CMRT+$E_c$ give the most accurate equilibrium properties in terms of $r_s$ and energy. However, they both underestimate the binding energy at $r_s \sim 1.3$ a.u..

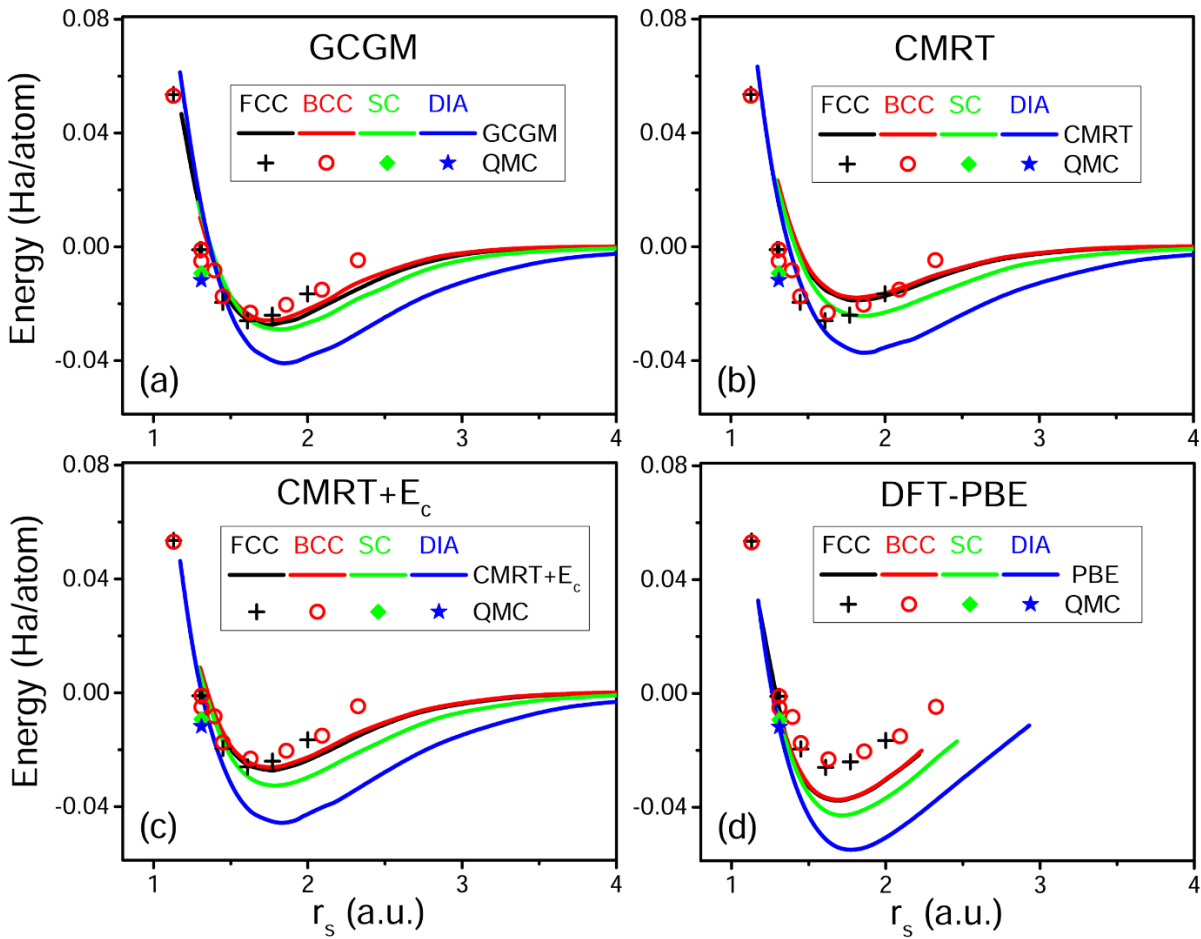

Figure 3. Equation of state for all the four phases given by DMC and (a) GCGM, (b) CMRT, (c) CMRT $+E_c$, and (d) DFT with the PBE functional.

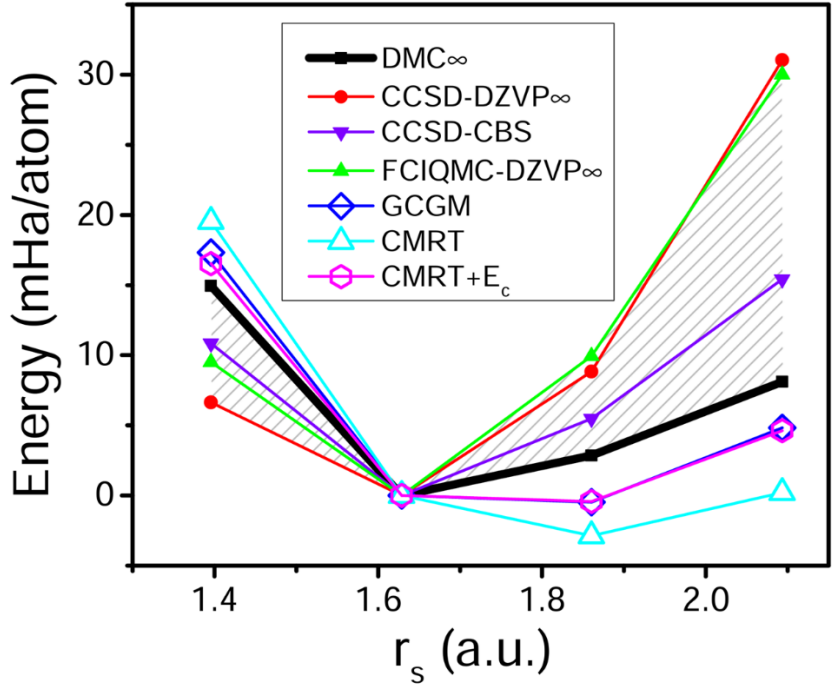

Figure 4. Equation of state (i.e., energy vs. $r_s$) for BCC-H obtained by GCGM, CMRT, CMRT$+E_c$ are compared with the results from other *ab initio* methods [37]. The energies are aligned together and shifted to 0 mHar/atom at $r_s = 1.6282$ a.u.. FCIQMC-DZVP∞ and CCSD-DZVP∞ denote estimated energies employing a DZVP basis set and extrapolated to the infinite system size limit. DMC∞ represents energies also extrapolated to the thermodynamic limit. CCSD-CBS energies are obtained using a finite simulation cell of 16 hydrogen atoms and are extropolated to the complete basis set (CBS) limit.

To assess the accuracy of our GCGM and CMRT methods against other ab initio methods currently used, in Fig. 4, we compare the equation of state for BCC-H in the range $1.4 \leq r_s \leq 2.2$ a.u. obtained by our GCGM, CMRT, CMRT$+E_c$ methods to the results from a variety of *ab initio* many-body wave function approaches, including DMC, coupled cluster singles and doubles (CCSD) [8,10,11] and full configuration interaction QMC (FCIQMC) [47] methods. The shaded region indicates the range of energies covered by DMC and the other reference methods. All three methods (FCIQMC, CCSD, and DMC), give very close equilibrium density for BCC-H at around $r_s = 1.65$ a.u.. Therefore, in Ref. [37] the energies from different methods are compared by shifting their energy minima to 0 mHar/atom at $r_s = 1.6282$ a.u.. Although GCGM and CMRT(+$E_c$) both give a bigger equilibrium density around $r_s = 1.72 \sim 1.76$ a.u., we also shift the energies from GCGM and CMRT(+$E_c$) to 0 at $r_s = 1.6282$ a.u., according to the way of data alignment in Ref. [37]. The results from GCGM and CMRT+$E_c$ are more closer to that of DMC as compared to the results from the other *ab initio* methods. CMRT without including $E_c$ yields more deviation from the DMC resulsts for $r_s > 1.6$ a.u.., but remains resonable accuracy.

The current calculations give ample information on how inclusion of residual correlation energy, $E_c$, into CMRT might benefit its accuracy in total energy calculations. However, $E_c$ in its current form adopted from LDA may not gurantee to improve the CMRT results for generic molecular or bulk systems. In this work, the inclusion of $E_c$ brings the CMRT results almost in line with GCGM for FCC and BCC lattices, but not for SC and DIA. Benchmarking the calculaton results from CMRT with those from GCGM are likely to provide guidance for developing a better and transferable expression for $E_c$ to consistently improve the CMRT energies for correlated materials.

Another alternative way to improve CMRT is by refining the form of the Gutzwiller-type approximation it employs. The current form of the Gutzwiller-type approximation used in CMRT, $\langle c_{i\alpha}^\dagger c_{j\beta} \rangle_{GWF} \approx z_{i\alpha} z_{j\beta} \langle c_{i\alpha}^\dagger c_{j\beta} \rangle_0$ with $z_{i\alpha} = \sqrt{z_{i\alpha}^{GA}}$, is determined from the requirement of CMRT reaching the exact ground state of a minimum basis Hydrogen dimer, $H_2$. This approximation is more accurate than the original Gutzwiller approximation in lower dimension but is still not accurate enough. GCGM can offer insights into developing a more accurate approximation for CMRT by comparing the density matrices produced by the two methods. In our future studies, we will extend GCGM to multi-band systems, compare the results from CMRT and GCGM calculations for systems with various degrees of correlations and various electron occupation fractions, and use the results to develop a more optimal form of $z_{i\alpha} = f\left(z_{i\alpha}^{GA}\right)$ in the approximation for CMRT.

Finally, we want to discuss the speed of GCGM and CMRT($+E_c$). It typically takes 3~8 core-hours for GCGM and 1~5 core-hours for CMRT to obtain the energies around equilibrium lattices for the four phases. The inclusion of $E_c$ in CMRT barely takes extra computational effort. Here, we use the SC-H phase as a case study to compare the computational efficiency of GCGM and CMRT with that of DMC. At equilibrium lattice ($r_s = 1.76$), both GCGM and CMRT use $16 \times 16 \times 16$ k-points, or equivalently, $4,096$ atoms. It takes approximately 6 core-hours for GCGM and 1.5 core-hours for CMRT to achieve energy convergence with gradient norm $< 10^{-5}$. We use the same QMC package named CASINO [48] as in Ref. [37] to test the speed of DMC. It takes ~330 core-hours for DMC to obtain energy with an uncertainty of 0.37 mHa/atom of a $6 \times 6 \times 6$ supercell with 216 atoms.

4. CONCLUSION

Although both GCGM and CMRT employ the Gutzwiller wavefunction, which has the form of a variational ansatz, the energy evaluation involves certain approximations introduced for computational efficiency. As a result, the strict variational principle no longer holds. This motivates the need for extensive benchmark tests to carefully assess the accuracy of our methods. In earlier work, we benchmarked our approaches against highly accurate reference results. Starting with the simplest cases, we have tested systems including molecular ground and excited states as well as one- and two-band Hubbard models in one and two dimensions [29-33]. The present study of atomic solid hydrogen phases represents a step beyond these bulk model systems, and we plan to extend our benchmarks to realistic multi-orbital bulk systems as our methods are further developed.

In this work, we compare our results with those given by DMC. The energy predicted by GCGM agrees reasonably well with the DMC data, especially for FCC- and BCC-H. Meanwhile, we present the results from CMRT, a method that is also based on GWF but uses a more simplified approximation to achieve an improved efficiency. As expected, the energy obtained by CMRT is less accurate than GCGM. Since both methods are developed within a similar scheme based on GWF, GCGM results can provide insights to improve the formalism of CMRT for better accuracy. The accuracy of CMRT strongly relies on the appropriate Gutzwiller-type approximations used in the theory. Since GCGM also uses GWF and yields more accurate results by not relying on the Gutzwiller-type approximation, the GCGM results can be used to develop a more optimal form of $z_{i\alpha} = f\left(z_{i\alpha}^{GA}\right)$ in the approximation for CMRT. An alternative way to improve CMRT is to include residual correlation energy, $E_c$, in the CMRT energy, which is shown to have positive effect on the accuracy of CMRT total energy by adopting the expression for $E_c$ from LDA. In the long term, we aim to refine this expression using GCGM results to achieve a consistent improvement of CMRT energies.

ACKNOWLEDGEMENTS

Work at Ames National Laboratory was supported by the U.S. Department of Energy (DOE), Office of Science, Basic Energy Sciences, Materials Science and Engineering Division including a grant of computer time at the National Energy Research Scientific Computing Center (NERSC) in Berkeley. Ames Laboratory is operated for the U.S. DOE by Iowa State University under Contract No. DE-AC02-07CH11358.